\documentclass[sigconf]{acmart}

\AtBeginDocument{%
  \providecommand\BibTeX{{%
    \normalfont B\kern-0.5em{\scshape i\kern-0.25em b}\kern-0.8em\TeX}}}

\copyrightyear{2022}
\acmYear{2022}
\setcopyright{acmcopyright}\acmConference[GLSVLSI '22]{Proceedings of the Great Lakes Symposium on VLSI 2022}{June 6--8, 2022}{Irvine, CA, USA}
\acmBooktitle{Proceedings of the Great Lakes Symposium on VLSI 2022 (GLSVLSI '22), June 6--8, 2022, Irvine, CA, USA}
\acmPrice{15.00}
\acmDOI{10.1145/3526241.3530341}
\acmISBN{978-1-4503-9322-5/22/06}

\usepackage{enumitem}

\usepackage{subcaption}        
\usepackage[version=4]{mhchem} 
\usepackage[ruled,vlined,linesnumbered]{algorithm2e} 
\newcommand\note[1]{\textcolor{brown}{#1}}          
\SetCommentSty{mycommfont}                          
\begin{document}

\title{Watermarked ReRAM: A Technique to Prevent Counterfeit Memory Chips}




\author{Farah Ferdaus, B. M. S. Bahar Talukder, and Md Tauhidur Rahman}
\affiliation{
  \institution{ECE Department, Florida International University}
  \city{Miami}
  \country{USA}
}
\email{{fferd006,bbaha007,mdtrahma}@fiu.edu}







\renewcommand{\shortauthors}{Ferdaus, et al.}

\begin{abstract}
\textit{Electronic counterfeiting} is a longstanding problem with adverse long-term effects for many sectors, remaining on the rise. This article presents a novel low-cost technique to embed watermarking in devices with resistive-RAM (ReRAM) by manipulating its analog physical characteristics through switching (\textit{set/reset}) operation to prevent counterfeiting. We develop a system-level framework to control memory cells’ physical properties for imprinting irreversible watermarks into commercial ReRAMs that will be retrieved by sensing the changes in cells’ physical properties. Experimental results show that our proposed ReRAM watermarking is robust against temperature variation and acceptably fast with ${\sim}0.6bit/min$ of imprinting and ${\sim}15.625bits/s$ of retrieval rates.
\end{abstract}

\ccsdesc[500]{Hardware~Memory and dense storage}
\ccsdesc[500]{Security and privacy~Security in hardware}

\keywords{Watermarking, ReRAM, Counterfeiting, Supply-chain Security}


\maketitle

\section{Introduction} \label{sec:intro}
Fabricating chips in untrusted facilities is increasing worldwide, which paves the way for an easy entrance of counterfeit chips into the supply chain in different formats, such as recycled, remarked or forged documentation, tampered, cloned, reverse-engineered, out-of-spec/defective, and overproduced \cite{Ioannis, guin, Forte:CHES, PVal:Basak, JV:Counterfeit,rahman2021systems,contreras2013secure}. Recent studies show that memory and memory integrated ICs (microprocessors, programmable logic devices, etc.) consist of ${\sim}$50\% of the total counterfeit market share \cite{Forte:CHES}. Most counterfeit memory chips suffer from sub-standard quality, poor performance, and shorter lifespan, severely affecting the security and reliability domains \cite{guin, Forte:CHES,talukder2020towards}. To date, there have been several anti-counterfeiting solutions to avoid fake chips, such as hardware metering, secured split testing (SST), on-chip sensor, split manufacturing, electronic chip ID, IC camouflaging, DNA marking, physical inspection-based test, burn-in test, and electrical test \cite{Forte:CHES, guin, PVal:Basak, JV:Counterfeit}. Unfortunately, all of these techniques suffer at least one of the following limitations- (i) focused on a single counterfeit type (e.g., only identifying remarked chips), (ii) requires hardware modification, (iii) involves complex supply chain management, (iv) requires help from the subject-matter of experts, (v) suffers from low test accuracy, and (vi) requires expensive lab facility \cite{guin, guin_comprehensive, Forte:CHES}. In contrast, watermarking is considered a cost-effective anti-counterfeit solution because watermark imprint/extraction can be performed without circuit modification, subject-matter experts, or extensive testing \cite{Watermarking}.

This article focuses on preventing counterfeit ReRAM chips or chips with embedded ReRAM by watermarking technique. The emerging ReRAM has several advantages: architectural simplicity, high scalability, ultra-low power operation, high density, cross-bar structure feasibility, excellent reliability at high temperature, high endurance compared to other traditional storage memories. \cite{ReRAM_Yi, ReRAM_Yang, Fujitsu}. Therefore, ReRAM has been investigated to a great extent to integrate into low-power applications, such as the Internet of Things (IoT), wearable devices (e.g., smartwatch, smart glasses), tablets, smartphones, automobiles, and medical devices (e.g., hearing aids). Such elevated use of ReRAMs makes it a lucrative target to counterfeiters. Our aim is to prevent counterfeiting of such chips by embedding watermarks in ReRAM cells by leveraging analog characteristics of ReRAM.

Technically, ReRAM is analogous to a two-terminal passive variable resistor where two resistance states, high resistance state ($HRS$) and low resistance state ($LRS$), represent the binary data values. Our technique imprints the watermark by repeatedly stressing the memory cells by alternatively writing ‘1’ and ‘0’. Repeated stressing through switching operation (`1’ $\rightarrow$ `0’ or `0’ $\rightarrow$ `1’) gradually decreases the $HRS$ resistance, degrading the memory performance and eventually causing endurance failure \cite{ReRAM_Switching, ReRAM_Mao}. Our experiment indicates that repeatedly stressing the ReRAM cell increases its \textit{write} time (for both logic `0' and `1'). To this extent, we propose a technique of imprinting logic `0' and `1' by representing the fresh and stressed memory cells, respectively. Later, we retrieve the imprinted sequence by observing the \textit{write} time of corresponding memory cells. Our proposed technique is irreversible as the impact of cell stressing is immutable. Hence, the imprinted watermark cannot be tampered. Additionally, our proposed technique does not require any hardware modification and can be directly deployed into available commercial products. Furthermore, the embedded watermark is robust against temperature variation as ReRAM is inherently insensitive to temperature \cite{Bogdan:ReRAM}. Moreover, our proposed method can be evaluated using standard ReRAM \textit{read/write} operation and only costs ${\sim}2$\% of the total endurance of ReRAM cells. The major contributions of this work are as follows.

\begin{itemize} [leftmargin=*, topsep=0pt,itemsep=-1ex,partopsep=1ex,parsep=1ex]
    \item We characterize the impact of repeated stressing on ReRAM \textit{write} time experimentally and show that the ReRAM \textit{write} time increases monotonically with respect to the stress count.
    \item We present a novel idea of ReRAM watermarking by storing logic `0' bit in fresh ReRAM cells and logic `1' in stressed ReRAM cells. We experimentally show that the imprinted data can be retrieved by observing ReRAM \textit{write} time.
    \item {We demonstrate the system throughput and verify the robustness of our proposed watermarking technique in multiple commercial off-the-shelf (COTS) ReRAM chips.}
\end{itemize}

The rest of the paper is organized as follows. Sec. \ref{sec:reram} briefly overviews the ReRAM memory preliminaries. Sec. \ref{sec:method} presents the proposed watermark imprinting and extracting mechanism, including the method for characterization of changes in ReRAM \textit{write} time caused by stress. Sec. \ref{sec:evaluate} explains the experimental setup and exhibits obtained results. Finally, Sec. \ref{sec:end} concludes our work.

\section{ReRAM Preliminaries} \label{sec:reram}
\begin{figure}[ht!]
    \centering
    \captionsetup{justification=centering, margin= 0cm}
    \includegraphics[trim=0cm 12.4cm 20.5cm 0.1cm, clip, width = 0.32\textwidth]{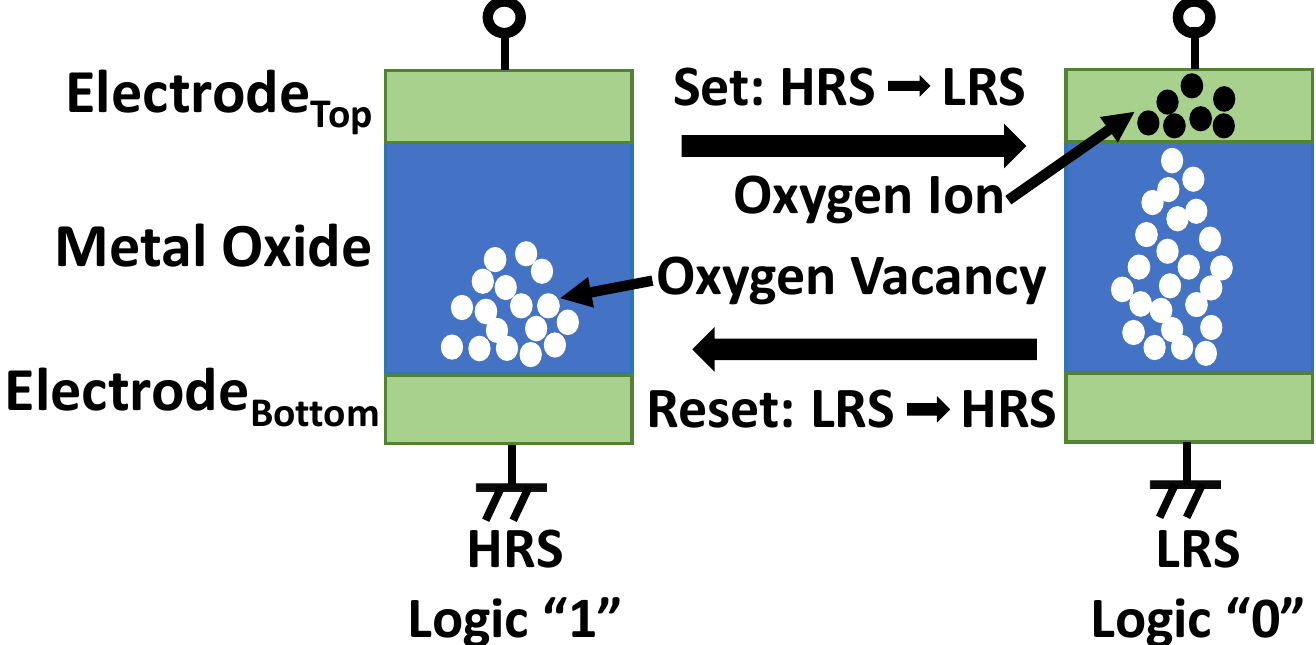}
    \caption{ReRAM cell structure with two logic states \cite{ReRAM_Mao}.}
    \label{fig:ReRAM}
\end{figure}

Resistive switching phenomena in a dielectric material is the core mechanism of ReRAM to store logic states \cite{ReRAM_hist, ReRAM_Mao}. The capacitor-like ReRAM bit cell structure consists of two electrodes ($Electrode_{Top}$ and $Electrode_{Bottom}$) separated by a metal oxide resistive switch material (Fig.~\ref{fig:ReRAM}). Studies show that various metal oxide materials can be used to build the resistive switch layer, such as  $\ce{Al2O3}$, $\ce{NiO}$, $\ce{SiO2}$, $\ce{Ta2O5}$, $\ce{ZrO2}$, $\ce{TiO2}$, $\ce{HfO2}$, and $\ce{Nb2O5}$ \cite{ReRAM_hist, ReRAM_Mao}. However, different materials result in different device characteristics such as endurance, retention, and scalability \cite{ReRAM_hist, ReRAM_Mao}. Whenever a voltage is applied to the $Electrode_{Top}$, the metal oxide breakdown process is initiated and produces oxygen vacancies in the oxide layer. Consequently, these oxygen vacancies form a conductive filament between two electrodes and produce the low resistance state ($LRS$ or logic `0' state). A voltage with opposite polarity is applied across the metal oxide to eliminate the conductive filament, representing the high resistance state ($HRS$ or logic `1' state) of the ReRAM cell. The ratio between $HRS$'s resistance to LRS's is required to be large enough to ensure robust \textit{read/write} operation \cite{ReRAM_Mao}. The switching operations from $HRS$ ($LRS$) to $LRS$ ($HRS$) is known as \textit{set} (\textit{reset}) operation, and the time required for switching is known as the \textit{set} (\textit{reset}) time. In summary, the ReRAM \textit{read/write} operation is performed as follows:
\begin{itemize} [leftmargin=*, topsep=0pt,itemsep=-1ex,partopsep=1ex,parsep=1ex]
    \item The \textit{write} operation ensures appropriate voltage magnitude and polarity across the ReRAM cell; as a result, the ReRAM cell obtains the appropriate resistance state ($LRS$ for logic `0' and $HRS$ for logic `1'). 
    \item During the \textit{read} operation, a small voltage is applied across the ReRAM bit cell, and the measured resistance (by sensing current) determines the stored logic state.
\end{itemize}

Each switching operation (i.e., changing state from $LRS$ to $HRS$ or $HRS$ to $LRS$) on ReRAM gradually decreases the resistance of $HRS$, wearing-out the device \cite{ReRAM_Switching}. Hence, fresh memory cells possess distinctly different analog properties from the stressed cells (i.e., cells that undergo repeated switching operations). For example, the reduction of resistance of $HRS$ due to the wear-out process degrades the resistance ratio of $HRS/LRS$ \cite{ReRAM_Switching, ReRAM_Mao}. To maintain the desired resistance ratio of $HRS/LRS$, \textit{set} and \textit{reset} times must be increased for stressed memory cells\footnote{The ReRAM internal control circuit maintains appropriate \textit{set/reset} time by initiating write-verify-write operation sequence \cite{jain:ReRAM}.}. In this work, we use this property to distinguish between the fresh and stressed ReRAM cells.

\section{Proposed Watermarking Technique} \label{sec:method}
\begin{figure}[ht!]
    \centering
    \captionsetup{justification=centering, margin= 0cm}
    \includegraphics[trim=0cm 11.6cm 19.5cm 0cm, clip, width = 0.3\textwidth]{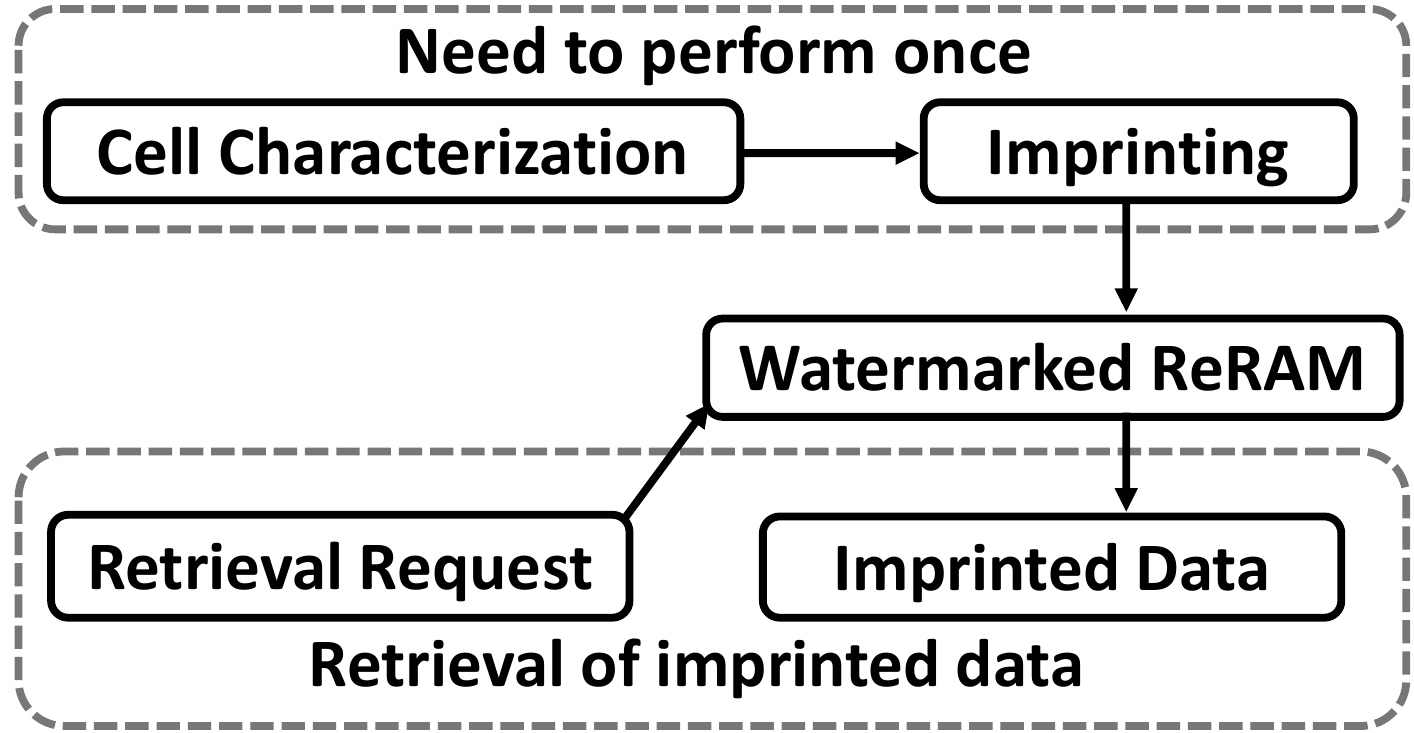}
    \caption{Steps used for ReRAM watermarking.}
    \label{fig:wMark_step}
\end{figure}

The flowchart in Fig. \ref{fig:wMark_step} shows the steps of imprinting watermark chronologically. At first, we characterize a few memory cells to understand the analog physical characteristics of ReRAM cells at different stressing levels up to the maximum endurance. Second, we imprint watermarks through repeated stressing the memory cells. These two steps are required to be performed only once. Finally, in the retrieval step, the end-user or manufacturer extracts the physical properties of the memory cells through standard digital interfaces.

\subsection{Cell Characterization} \label{subsec:Char}
Repeated switching operations (alternatively writing 0's and 1's) change the physical properties of ReRAM; therefore, the \textit{set/reset} timing of stressed cells deviates from the fresh cells. The degree of deviation depends on the number of switching operations performed on stressed cells. Our proposed technique imprints logic `1’ with stressed cells and `0’ with fresh cells.  Later, we retrieve the data by separating the fresh cells and stressed cells based on their switching time. However, ReRAM stressing reduces cell endurance. Therefore, we want to keep the stress level as little as possible and simultaneously ensure that fresh and stressed cells are reliably separable with \textit{set/reset} time.

\begin{algorithm}[ht!]
\SetAlgoLined
    \KwData{
        $\mathcal{N_M}$: \note{Max rewrite operations (data endurance)} \break 
        $\mathcal{A_S}$: \note{Set of memory addresses targeted to stress} \break
        $w_L$: \note{Word length} \break
        $\mathcal{D}$:  \note{Data vector of length $w_L$, intended to write in target memory cells belong to $\mathcal{A_S}$} \break
        $t$: \note{Timer}} 
            
    \KwResult{
        $\mathcal{S_T}$: \note{\textit{Set} time of memory cells belongs to $\mathcal{A_S}$} \break
        $\mathcal{R_T}$: \note{\textit{Reset} time of memory cells belongs to $\mathcal{A_S}$}}
    \BlankLine    
    \tcp{Initialization} 
    $\mathcal{S_T} = \{ \};\ \mathcal{R_T} = \{ \};\ \mathcal{D} = Ones(1 \times w_L)$\;
    \ForEach{a $\in \mathcal{A_S}$}{ \label{alg1:init0}
        $write(a, \mathcal{D})$\;
    } \label{alg1:init1}
    
    \BlankLine
    \tcp{Stressing memory cells}
    \For{$i = 0$ to $\mathcal{N_M}$}{ \label{alg1:stress0}
        \ForEach{a $\in \mathcal{A_S}$}{
            $\mathcal{D} = Zeros(1 \times w_L)$\;
            $tic = t$\;
            $write(a, \mathcal{D});$ \tcp{\textit{Set} operation} \break
            $toc = t - tic$\;
            $\mathcal{S_T} = \mathcal{S_T} \cup \{toc\};$ \tcp{Accumulating \textit{set} time} \break 
            \BlankLine
            $\mathcal{D} = Ones(1 \times w_L)$\; 
            $tic = t$\;
            $write(a, \mathcal{D});$ \tcp{\textit{Reset} operation} \break
            $toc = t - tic$\;
            $\mathcal{R_T} = \mathcal{R_T} \cup \{toc\};$ \tcp{Accumulating \textit{reset} time} \break 
         }
    } \label{alg1:stress1}
 \caption{Pseudo-code for characterizing memory cells using repeated switching operation.}
 \label{alg:Char}
\end{algorithm}

To this extent, we propose Algorithm \ref{alg:Char} to understand the ReRAM cell characteristics and the impact of switching operation on \textit{set/reset} timing. This algorithm allows us to determine the minimum number of switching operations required to separate the stressed cell from the fresh cell reliably. It also builds a relationship between ReRAM switching time and corresponding stressing level. The sequence of operations for this algorithm is as follows. We initiate our algorithm by writing all `1’ data patterns to selected memory addresses (line \ref{alg1:init0} through line \ref{alg1:init1} of Algorithm \ref{alg:Char}). Then, all `0' and all `1' data patterns are written alternatively to those addresses (line \ref{alg1:stress0} through line \ref{alg1:stress1} of Algorithm \ref{alg:Char}). The switching times are captured and stored as \textit{set/reset} times accordingly. We repeat the switching operation until the target memory cells are fully worn-out (i.e., no longer able to store data reliably). We observe that both the \textit{set} and \textit{reset} times increase due to the repeated switching operation, and after a certain number of switching operations, the stressed cells completely become separable from fresh cells.

Note that, according to our observation, the relation between switching characteristics (i.e., \textit{set/reset} time vs. stress count\footnote{One `stress' means a pair of \textit{set-reset} operation.}) is almost uniform for all memory chips sharing the same part-number. Therefore, it should be sufficient to sample a small set of memory chips from each part-number and perform cell characterization over those chips.

\subsection{Imprinting Scheme} \label{subsec:imprint}
After characterization, our next step is to imprint watermarks in ReRAM. Chip manufacturers perform the proposed watermark imprinting technique into the memory during the die-sort testing phase \cite{sakib:flashmark}. The watermark may include standard device ID, chip-specific unique ID, and other manufacturing-related information \cite{sakib:flashmark}. In the proposed technique, we reserve a set of addresses for the watermark; the number of addresses depends on the length of the watermark. Initially, all memory cells possess perfect or near-perfect analog properties since they are fresh. To imprint watermarks, (i) initially, logic `1' is written to those reserved addresses (line \ref{alg2:init0} through line \ref{alg2:init1} of Algorithm \ref{alg:wMark}), and (ii) repeated switching (\textit{set} and \textit{reset}) operations are performed (line \ref{alg2:wmark0} through line \ref{alg2:wmark1} of Algorithm \ref{alg:wMark}) to only those ReRAM addresses, which are supposed to hold the logic `1' of target watermark. The switching operations are repeated until sufficient differences are developed in the \textit{set/reset} time between fresh cells and stressed memory cells. Each switching operation gradually degrades the resistance of $HRS$, which are permanent; thus cannot be reversed. However, the number of repeated switching cycles, $\mathcal{N}$, used to imprint the watermark must be determined through the cell characterization phase for given memory chips (see Sec. \ref{subsec:Char}). From an imprinting perspective, it is desirable to minimize $\mathcal{N}$ because the imprinting time of the watermark is directly proportional to the number of switching cycles. However, higher $\mathcal{N}$ enhances the accuracy by distinguishing fresh and stressed memory cells more perfectly.

\begin{algorithm}[ht!]
\SetAlgoLined
    \KwData{
        $\mathcal{N}$: \note{Number of stress count (i.e. \textit{set-reset} pairs)} \break 
        $\mathcal{A_W}$: \note{Set of memory addresses containing watermark.} \break
        $w_L$: \note{Word length} \break
        $wMark$: \note{Watermark} \break
        $\mathcal{D}$: \note{Data vector of length $w_L$, intended to write in target memory cells belong to $\mathcal{A_W}$} \break
        $t$: \note{Timer}} \break
            
    \KwResult{
        $\mathcal{S_T}$: \note{\textit{Set} time of memory cells belongs to $\mathcal{A_W}$} \break
        $\mathcal{R_T}$: \note{\textit{Reset} time of memory cells belongs to $\mathcal{A_W}$}}
    
    \BlankLine    
    \tcp{Initialization} 
    $\mathcal{S_T} = \{ \};\ \mathcal{R_T} = \{ \};\ \mathcal{D} = Ones(1 \times w_L)$\;
    \ForEach{a $\in \mathcal{A_W}$}{ \label{alg2:init0}
        $write(a, \mathcal{D})$\;
    } \label{alg2:init1}
    
    \BlankLine
    \tcp{Imprinting watermark}
    \For{$i = 0$ to $\mathcal{N}$}{ \label{alg2:wmark0}
        \ForEach{a $\in \mathcal{A_W}$}{ 
            \If{wMark[Bit]==1}{
                $\mathcal{D} = Zeros(1 \times wS)$\;
                $write(a, \mathcal{D})$\;
                $\mathcal{D} = Ones(1 \times wS)$\;
                $write(a, \mathcal{D})$\;
            } 
        }
    } \label{alg2:wmark1}
    
    \BlankLine
    \tcp{Extracting watermark}
    \ForEach{a $\in \mathcal{A_W}$}{ \label{alg2:read0}
        $\mathcal{D} = Zeros(1 \times w_L)$\;
        $tic = t$\;
        $write(a, \mathcal{D})$; \tcp{\textit{Set} operation} \break
        $toc = t - tic$; \tcp{Accumulating \textit{Set} time} \break
        $\mathcal{S_T} = \mathcal{S_T} \cup \{toc\}$\; 
        \BlankLine
        $\mathcal{D} = Ones(1 \times w_L)$\;
        $tic = t$\;
        $write(a, \mathcal{D})$; \tcp{\textit{Reset} operation} \break
        $toc = t - tic$; \tcp{Accumulating \textit{Reset} time} \break 
        $\mathcal{R_T} = \mathcal{R_T} \cup \{toc\}$\;
    }\label{alg2:read1}
 \caption{Pseudo-code for imprinting and extracting watermarks.}
 \label{alg:wMark}
\end{algorithm}

\subsection{Retrieval Scheme} \label{subsec:retrive}
System designers read watermarks to verify the chips' authenticity before incorporating them into the products or verify later in the product life-cycle. In order to retrieve watermarks and imprinted status information, the physical properties of memory cells are extracted (in our case, \textit{set/reset} times) to distinguish between fresh and stressed memory cells. Line \ref{alg2:read0} to \ref{alg2:read1} of Algorithm \ref{alg:wMark} outlines the required  steps of extracting the \textit{set} and \textit{reset} times from the watermarked addresses. We observe that both \textit{set} and \textit{reset} time change with stress counts, and both can be used to imprint watermarks. For example, the manufacturer can define a threshold value of \textit{set/reset} time after imprinting the watermark, which can be used to differentiate between fresh and stressed memory cells.

It is worth mentioning that \textit{set/reset} characteristics of ReRAM cells appear to be uniform across all ReRAM chips that we have tested. Therefore, the manufacturer can define a fixed standard set of addresses for all memory chips for watermarking. Such arrangement should simplify the evaluation process. For example, the manufacturer can make the addresses that are used for watermarking publicly available. Anyone with this information should be able to access the watermark data and verify the chip authenticity. 

\section{Results and Discussion} \label{sec:evaluate}
\subsection{Evaluation Setup and Analysis}\label{subsec:setup}
\begin{figure}[ht!]
    \centering
    \captionsetup{justification=centering, margin= 0.5cm}
    \begin{subfigure}[t]{0.235\textwidth}
        \centering
        \includegraphics[trim=0.2cm 0.3cm 0.2cm 0.5cm, clip, width = 0.9\textwidth]{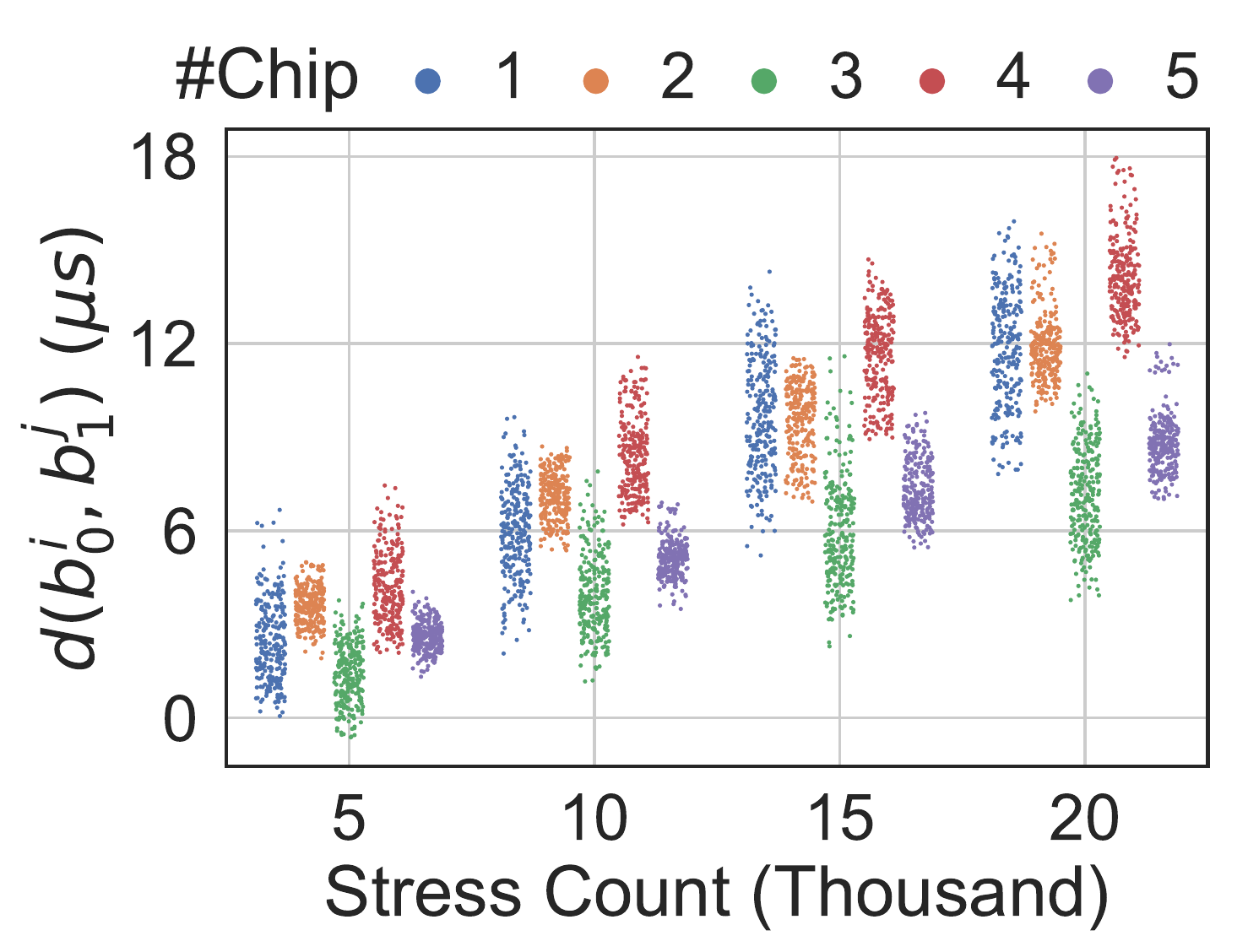}
        \caption{}
        \label{fig:summary_Set}
    \end{subfigure}
    \begin{subfigure}[t]{0.235\textwidth}
        \centering
        \includegraphics[trim=0.3cm 0.3cm 0.4cm 0.5cm, clip, width = 0.9\textwidth]{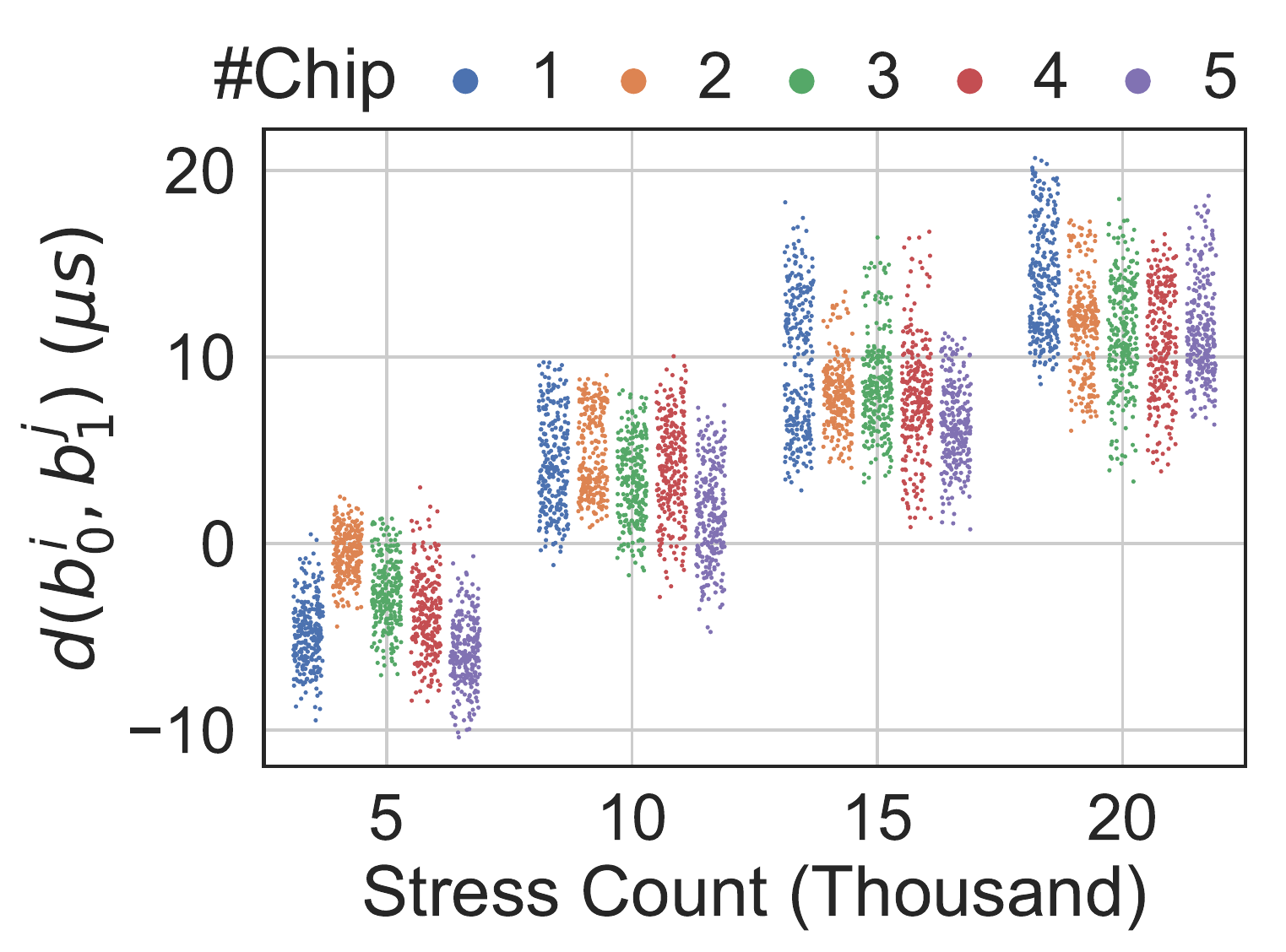}
        \caption{}
        \label{fig:summary_Reset}
    \end{subfigure}
    \caption{Verifying watermark in test chips, using- (a) $t_{Set,256}$, and (b) $t_{Reset,256}$.}
    \label{fig:summary}
\end{figure}

The analysis is performed over five \textit{MB85AS8MT}\footnote{We have also verified our proposed technique with \textit{MB85AS4MT} ReRAM chips produced by the same manufacturer. However, the Fujitsu \textit{MB85AS4MT} (180nm technology node) ReRAM chip is commercially discontinued, and the \textit{read/write} operation is much slower than the \textit{MB85AS8MT}. If the reviewers want, we will present data for \textit{MB85AS4MT} chips as well.} (40nm technology node) 8-bit serial peripheral interfaced (SPI) $8Mb$ memory chips manufactured by Fujitsu Semiconductor Limited. We have used our own custom-designed memory controller implemented on \textit{Teensy 4.1} microcontroller development board. 
The \textit{MB85AS8MT} ReRAM chips are byte-addressable. Therefore, a single byte is the smallest unit for which we can measure \textit{set/reset} time. As a result, we need at least a one-byte storage area in the ReRAM to imprint a single bit of data. However, the measured \textit{set/reset} time might vary due to the external and internal noise. Therefore, we imprint a single bit data into 256 consecutive addresses of the ReRAM to suppress the impact of noise. During evaluation, we have measured \textit{set/reset} time for each address and computed the average. From now on to the rest of the paper, we denote the average \textit{set/reset} time over 256 addresses as $t_{Set,256}$, and $t_{Reset,256}$, respectively. Note that the \textit{write buffer} size of our tested ReRAMs is also 256, which enables us to stress 256 addresses with a single \textit{write} command and hence, reduces overall stressing time. Although the figures (except Fig. \ref{fig:summary}) we present in this section are based on one ReRAM chip (randomly chosen from five test chips), the observation is valid for all test chips. Additionally, the Fig. \ref{fig:summary} summarizes the result from all five test chips.

\begin{figure}[ht!]
    \centering
    \captionsetup{justification=centering, margin= 0.5cm}
    \begin{subfigure}[t]{0.49\textwidth}
        \centering
        \includegraphics[trim=0.2cm 9cm 4.5cm 0.1cm, clip, width=0.9\textwidth]{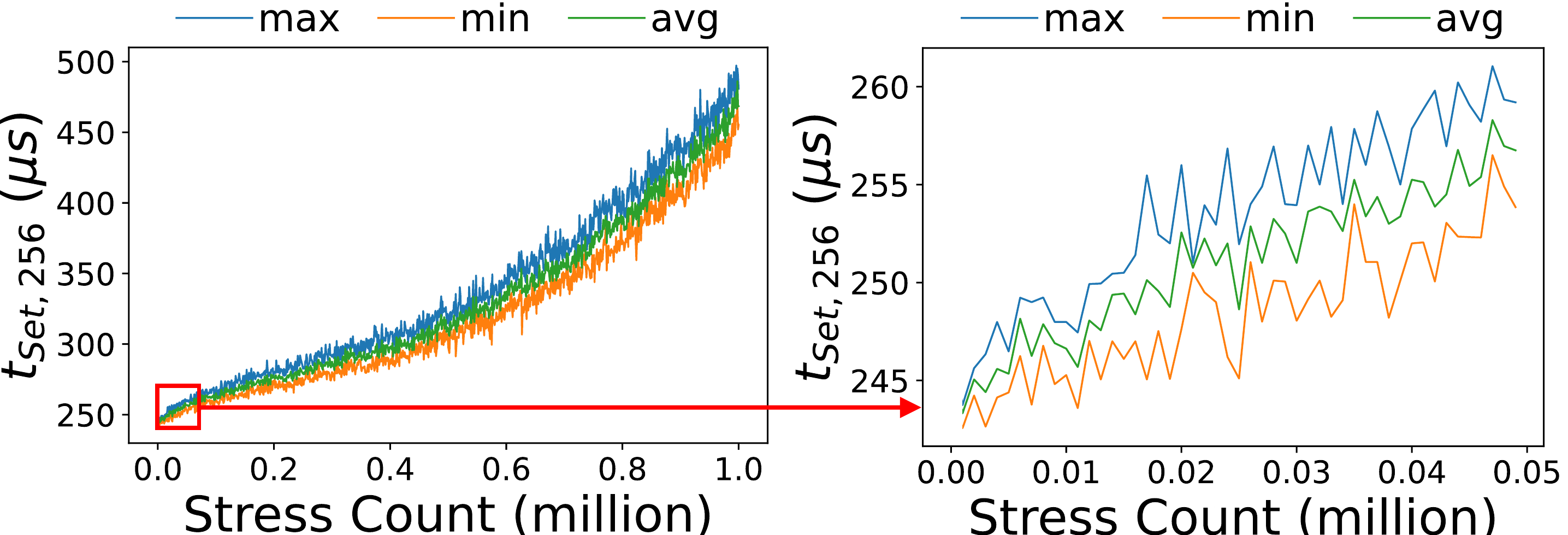}
        \caption{}
        \label{fig:SetT_char}
    \end{subfigure}%
    \vspace{\medskipamount}
    \begin{subfigure}[t]{0.485\textwidth}
        \centering
        \includegraphics[trim=0.2cm 8.9cm 4.4cm 0cm, clip, width = 0.9\textwidth]{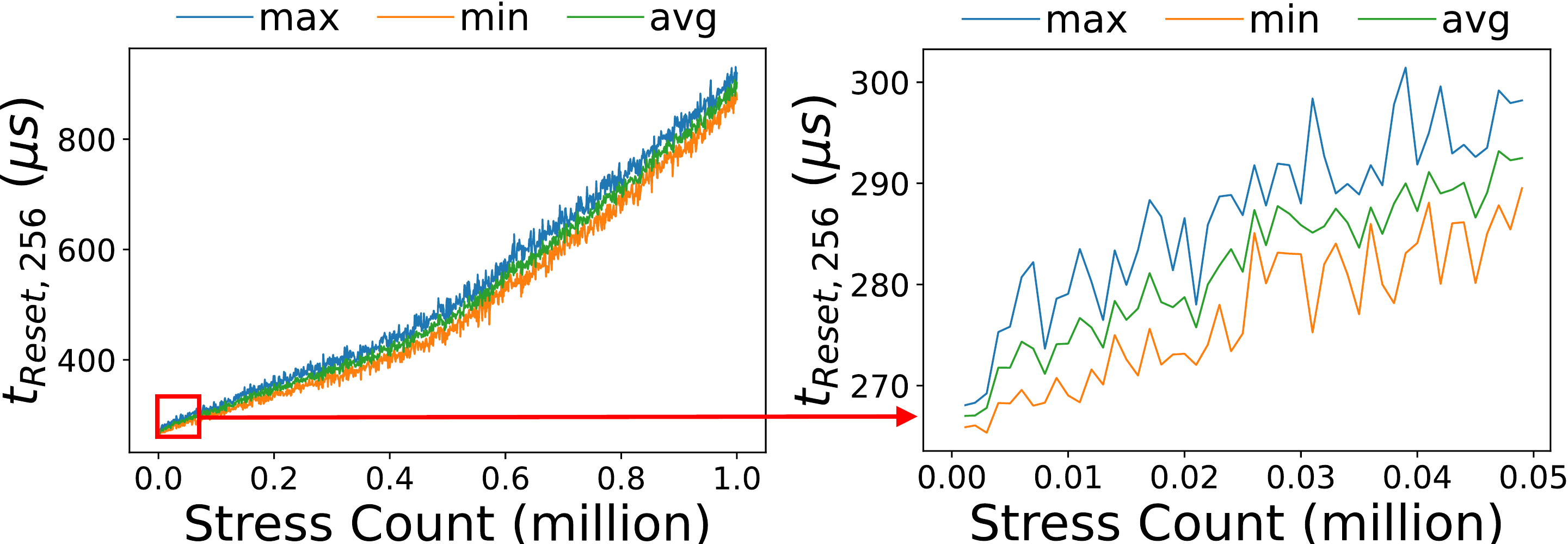}
        \caption{}
        \label{fig:ResetT_char}
    \end{subfigure}
    \caption{ReRAM cell characterization under stress- (a) $t_{Set,256}$ and (b) $t_{Reset,256}$.}
    \label{fig:char}
\end{figure}

Fig. \ref{fig:char} shows the switching characteristics (\textit{set/reset} time vs. the stress counts) of the ReRAM chips at $25^{\circ}C$. This figure represents the maximum, minimum, and average of $t_{Set,256}$ (Fig. \ref{fig:SetT_char}) and $t_{Reset,256}$ (Fig. \ref{fig:ResetT_char}) as a function of different stress levels (up to maximum possible rewrite operations\footnote{Maximum rated endurance for \textit{MB85AS8MT} ReRAM chip is $1M$ rewrite cycles (i.e., $500K$ \textit{set-reset} pairs). However, we observe that most memory cells can endure more rewrite operations than the rated endurance. In our experiment, we stress memory cells with up to $1M$ \textit{set-reset} pairs.}) over the 2K random address-space. Fig. \ref{fig:char} demonstrates that both the $t_{Set,256}$ and $t_{Reset,256}$ increase monotonically with stress levels, making it possible to distinguish between stressed and fresh memory cells. For example, the right-side zoomed plot of Fig. \ref{fig:SetT_char}, and \ref{fig:ResetT_char} represents \textit{set/reset} time up to $50K$ stress count, which demonstrates that the minimum value of $t_{Set,256}$ and $t_{Reset,256}$ at stressed count ${\sim}12K$ is larger than the maximum value of $t_{Set,256}$ and $t_{Reset,256}$ at fresh condition. Therefore, a proper threshold value of $t_{Set,256}$ or $t_{Reset,256}$ can reliably identify fresh cells and stressed cells with ${\sim}12K$ \textit{set/reset} operations. Although Fig. \ref{fig:char} is constructed with $2K$ memory addresses, a similar characteristic is valid for the whole address space.

Next, the following steps are performed to verify the feasibility of the proposed watermarking. We have imprinted an arbitrarily chosen 32-bit random data into $(256 \times 32) = 8192$ memory addresses varying the number of switching cycles, $\mathcal{N}$, up to $20K$ times to experimentally demonstrate the watermark imprinting (discussed in Sec. \ref{subsec:imprint}) and retrieval (discussed in Sec. \ref{subsec:retrive}) process. 

\begin{figure}[ht!]
\centering
\captionsetup{justification=centering, margin= 0cm}
\begin{minipage} []{.235\textwidth}
    \centering
    \captionsetup{justification=centering, margin= 0cm}
    \begin{subfigure}[t]{1\textwidth}
        \centering
        \includegraphics[trim=0cm 0cm 0cm 0cm, clip, width=0.9\textwidth]{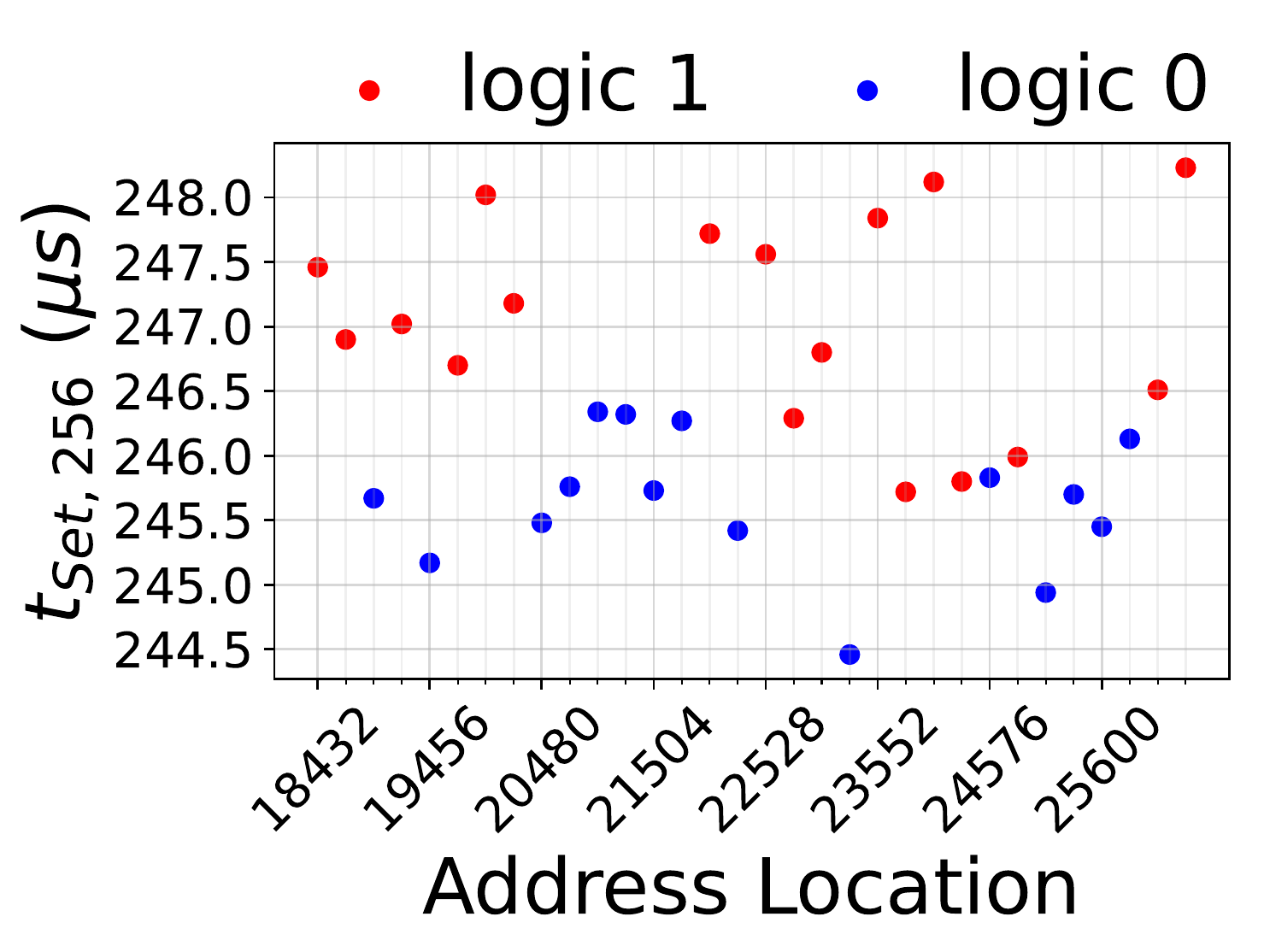}
        \caption{$\mathcal{N}=5K$}
        \label{fig:SetT_rep_5K}
    \end{subfigure}%
    \vspace{\medskipamount}
    \begin{subfigure}[t]{1\textwidth}
        \centering
        \includegraphics[trim=0cm 0cm 0cm 0cm, clip, width = 0.9\textwidth]{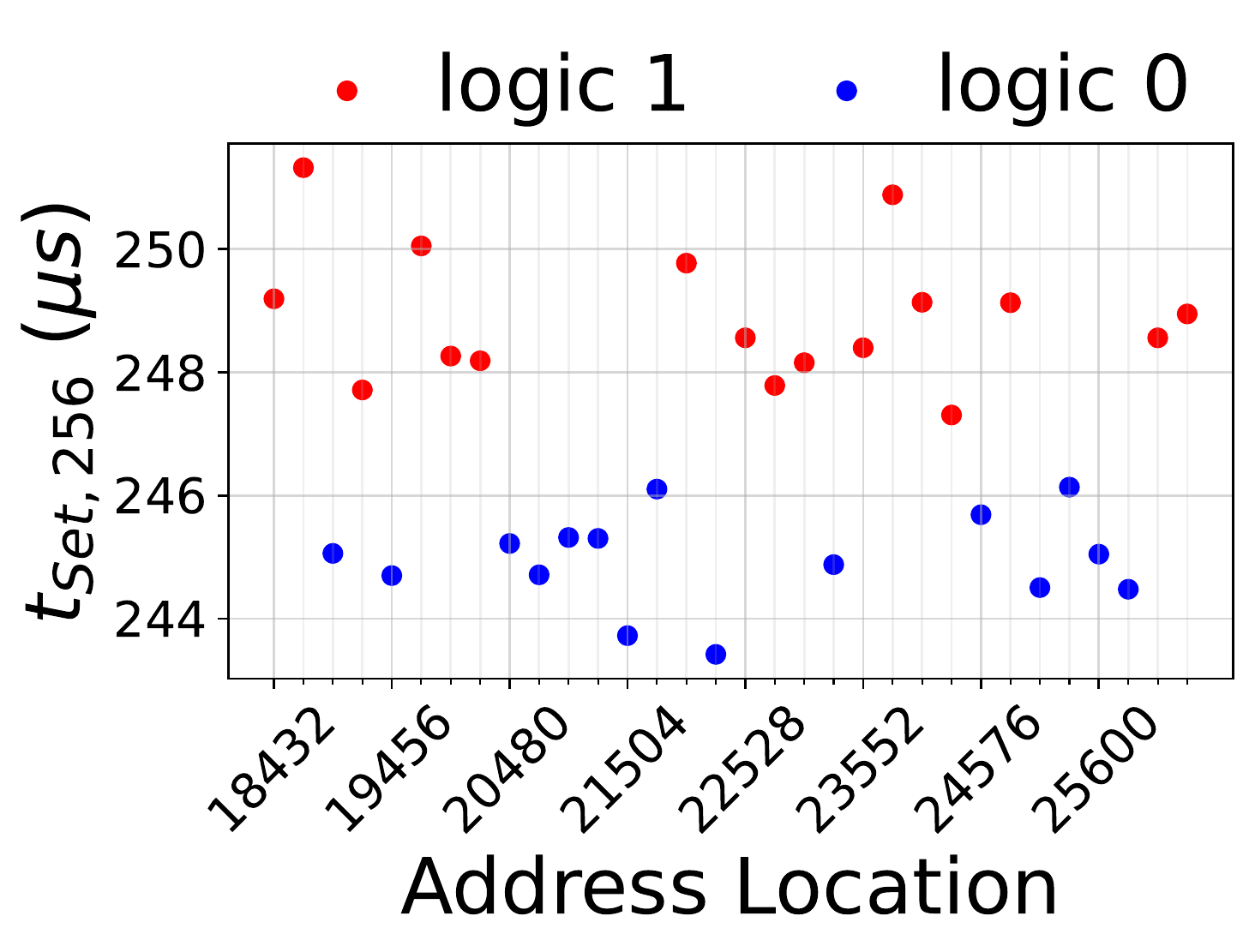}
        \caption{$\mathcal{N}=10K$}
        \label{fig:SetT_rep_10K}
    \end{subfigure}%
    \vspace{\medskipamount}
    \begin{subfigure}[t]{1\textwidth}
        \centering
        \includegraphics[trim=0cm 0cm 0cm 0cm, clip, width = 0.9\textwidth]{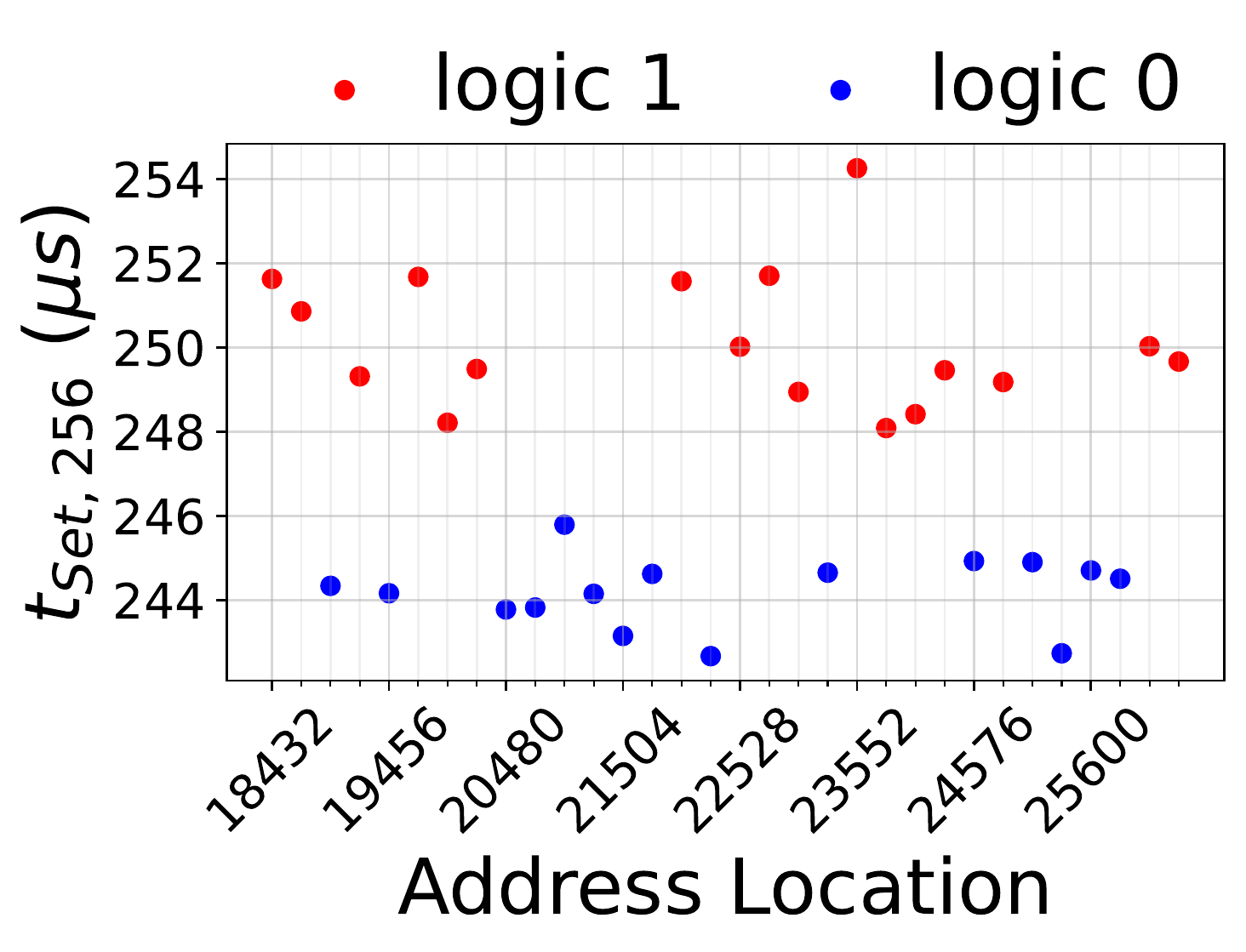}
        \caption{$\mathcal{N}=15K$}
        \label{fig:SetT_rep_15K}
    \end{subfigure}
\end{minipage}
\begin{minipage} []{.235\textwidth}
    \centering
    \captionsetup{justification=centering, margin= 0cm}
    \begin{subfigure}[t]{1\textwidth}
        \centering
        \includegraphics[trim=0cm 0cm 0cm 0cm, clip, width=0.9\textwidth]{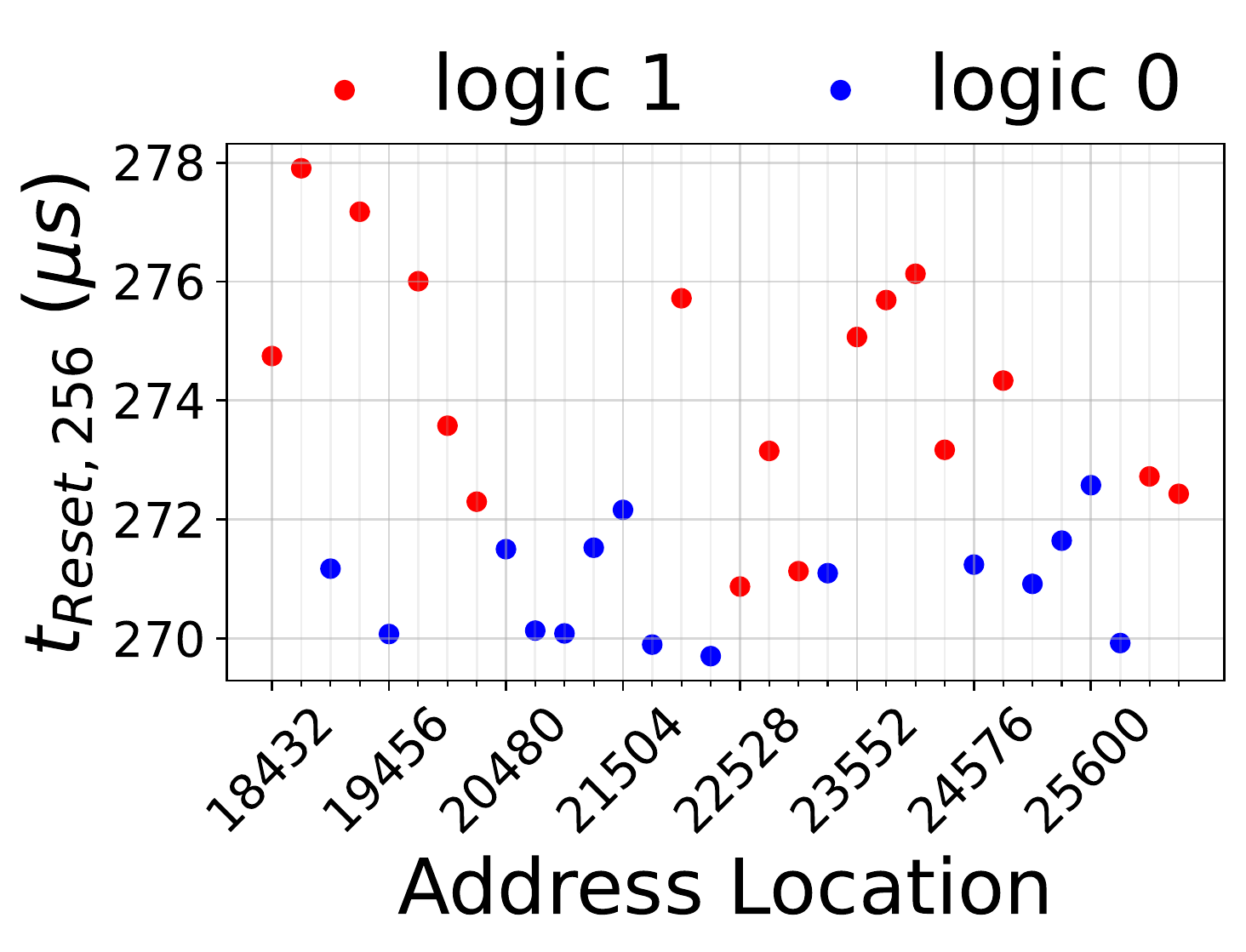}
        \caption{$\mathcal{N}=10K$}
        \label{fig:ResetT_rep_10K}
    \end{subfigure}%
    \vspace{\medskipamount}
    \begin{subfigure}[t]{1\textwidth}
        \centering
        \includegraphics[trim=0cm 0cm 0cm 0cm, clip, width = 0.9\textwidth]{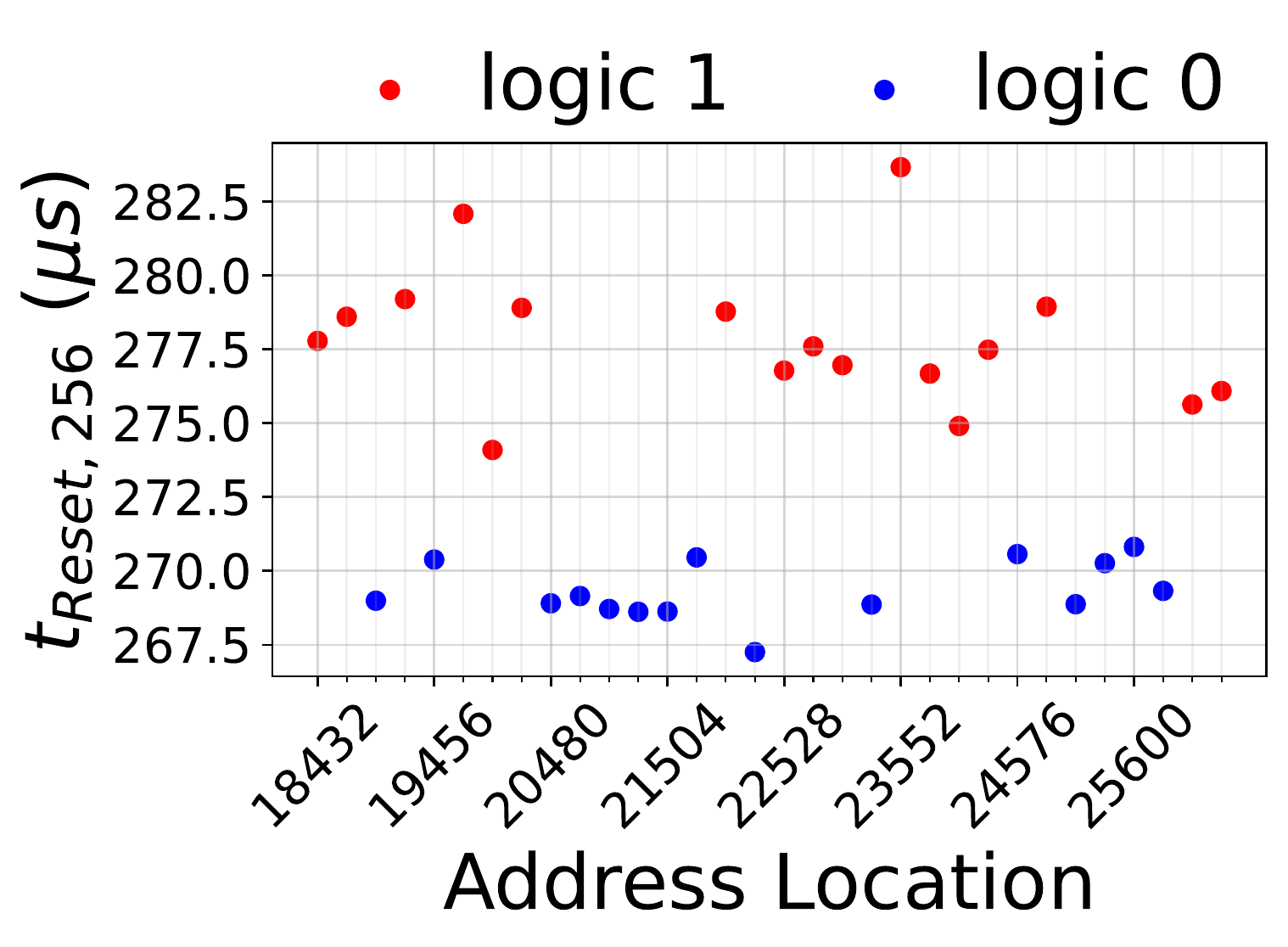}
        \caption{$\mathcal{N}=15K$}
        \label{fig:ResetT_rep_15K}
    \end{subfigure}%
    \vspace{\medskipamount}
    \begin{subfigure}[t]{1\textwidth}
        \centering
        \includegraphics[trim=0cm 0cm 0cm 0cm, clip, width = 0.9\textwidth]{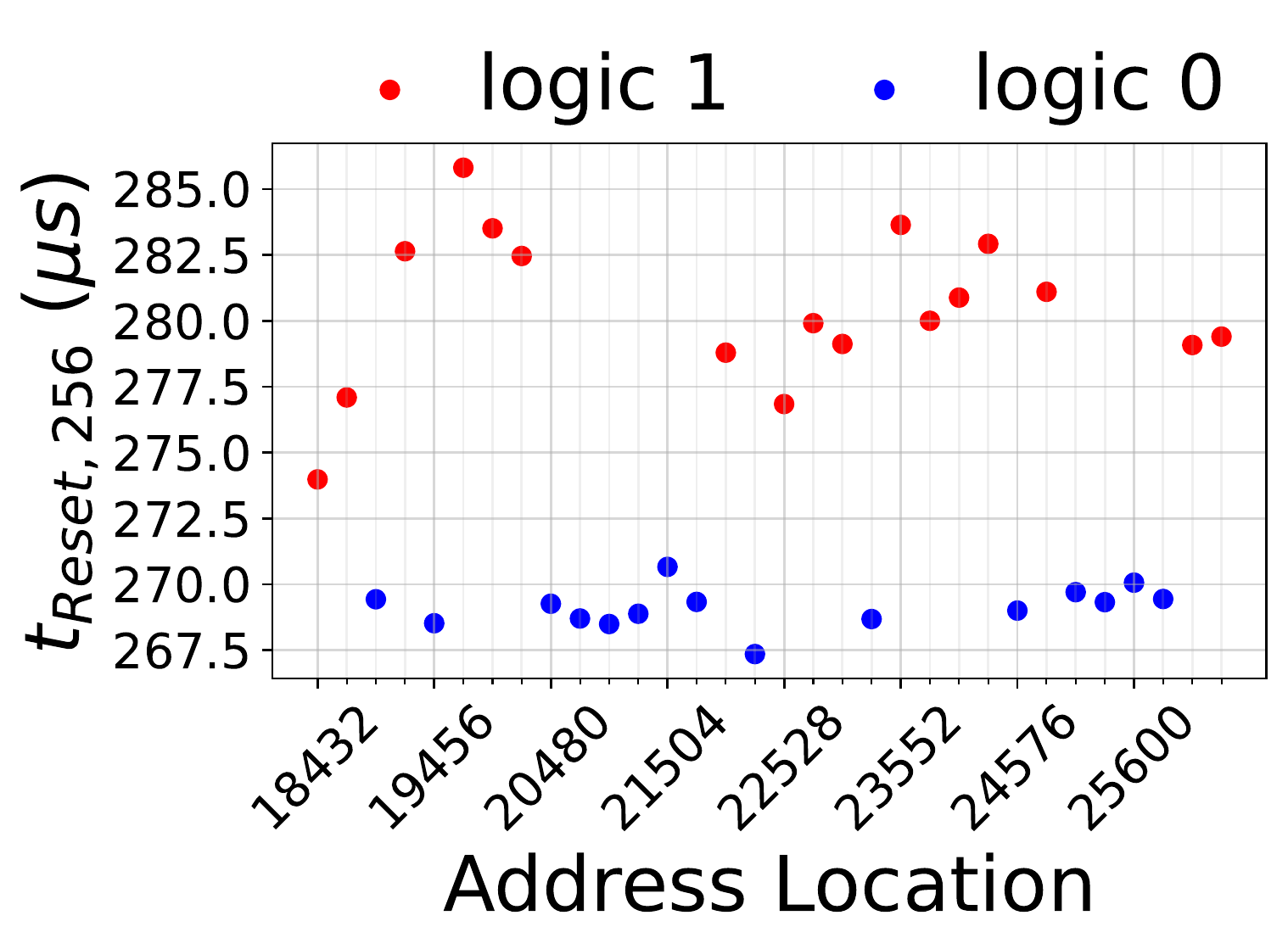}
        \caption{$\mathcal{N}=20K$}
        \label{fig:ResetT_rep_20K}
    \end{subfigure}
\end{minipage}
\caption{Imprinted data at different stress count-\\ (a)--(c) $t_{Set,256}$ at stress count $5K$, $10K$, and $15K$; \\ (d)--(f) $t_{Reset,256}$ at stress count $10K$, $15K$, and $20K$.}
\label{fig:T_rep}
\end{figure}

Fig. \ref{fig:T_rep} represents the experimental data from arbitrarily chosen test chips with imprinted data 0xC2F740EB\footnote{Also verified for other random data.}. We imprint the data in a random memory location. The red and blue dot represents the imprinted logic 1's and 0's, respectively. Fig. \ref{fig:T_rep} shows that logic `1' and logic `0' begin to separate at $5K$ stress count (Fig. \ref{fig:SetT_rep_5K}), and they become well-separated at $10K$ stress count (Fig. \ref{fig:SetT_rep_10K}). With further stress, the separation between logic `1' and logic `0' further increases (Fig. \ref{fig:SetT_rep_15K}). Similarly, with $t_{Reset,256}$, logic `1' and logic `0' begin to separate at $10K$ stress count (Fig. \ref{fig:ResetT_rep_10K}) and become well-separated at $15K$ stress count (Fig. \ref{fig:ResetT_rep_10K}). Therefore, with a proper threshold value of $t_{Set,256}$ (at $10K$ stress) or $t_{Reset,256}$ (at $15K$ stress), one can easily separate logic `0' and logic `1' bits.

Fig. \ref{fig:summary} verifies the watermark data imprinted in all five test memory chips. This figure represents the distribution of $d(b_0,b_1)$ at a different level of stresses, where $d(b_0,b_1)$ represents the distance between logic `0' bits ($b_0$) and logic `1' bits ($b_1$). Each dot in Fig. \ref{fig:summary} represents $d(b_0^i,b_1^j)$ for each possible $(b_0^i,b_1^j)$. For well-separated logic `0' and `1', the distance should be positive. A larger value of $d(b_0^i,b_1^j)$ is more desirable as it provides better separation between logic `0' and logic `1' bits. However, if the maximum value of \textit{set/reset} time of logic `0' bits is larger than the minimum value of \textit{set/reset} time of logic `1' bits (similar to Fig. \ref{fig:SetT_rep_5K}), then logic `0' bits and logic `1' bits cannot be separated properly. In such a scenario, the $d(b_0^i,b_1^j)$ can be negative for a few pairs of $(b_0^i,b_1^j)$. The figure demonstrates that the separation between logic `0' bits and logic `1' bits improves monotonically with respect to stress count. For all test chips, the logic `0' bits ($b_0$) and logic `1' bits ($b_1$) are clearly separable after $10K$ stresses with $t_{Set,256}$ and $15K$ stresses with $t_{Reset,256}$ (i.e., $\min \big( d(b_0^i,b_1^j) \big) > 0$).

\subsection{Robustness Analysis}\label{subsec:robust}
\begin{figure}[htp]
    \centering
    \captionsetup{justification=centering, margin= 0.5cm}
    \begin{subfigure}[t]{0.235\textwidth}
        \centering
        \includegraphics[trim=0cm 0cm 0cm 0cm, clip, width = 0.9\textwidth]{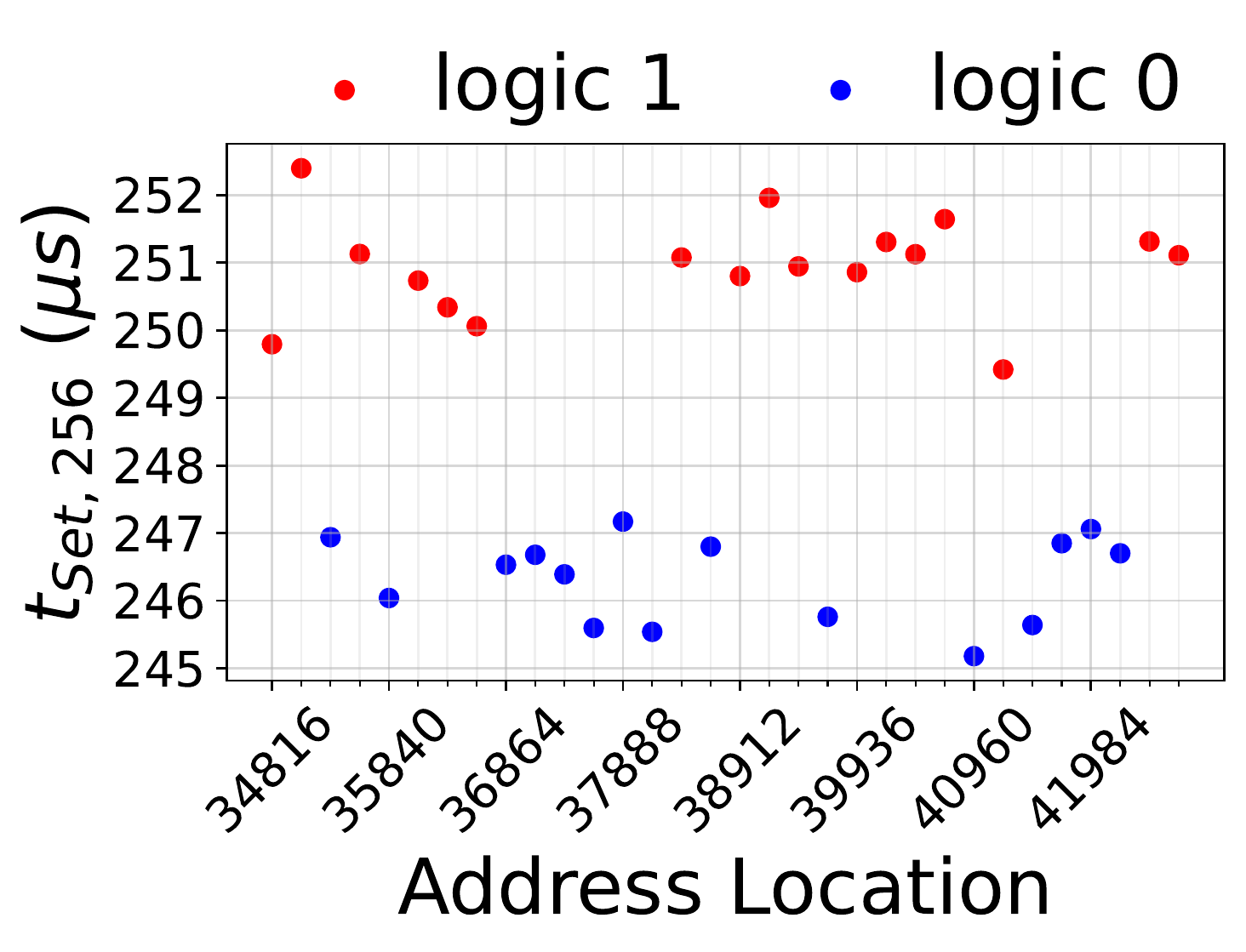}
        \caption{}
        \label{fig:bake_HT}
    \end{subfigure}
    \begin{subfigure}[t]{0.235\textwidth}
        \centering
        \includegraphics[trim=0cm 0cm 0cm 0cm, clip, width = 0.9\textwidth]{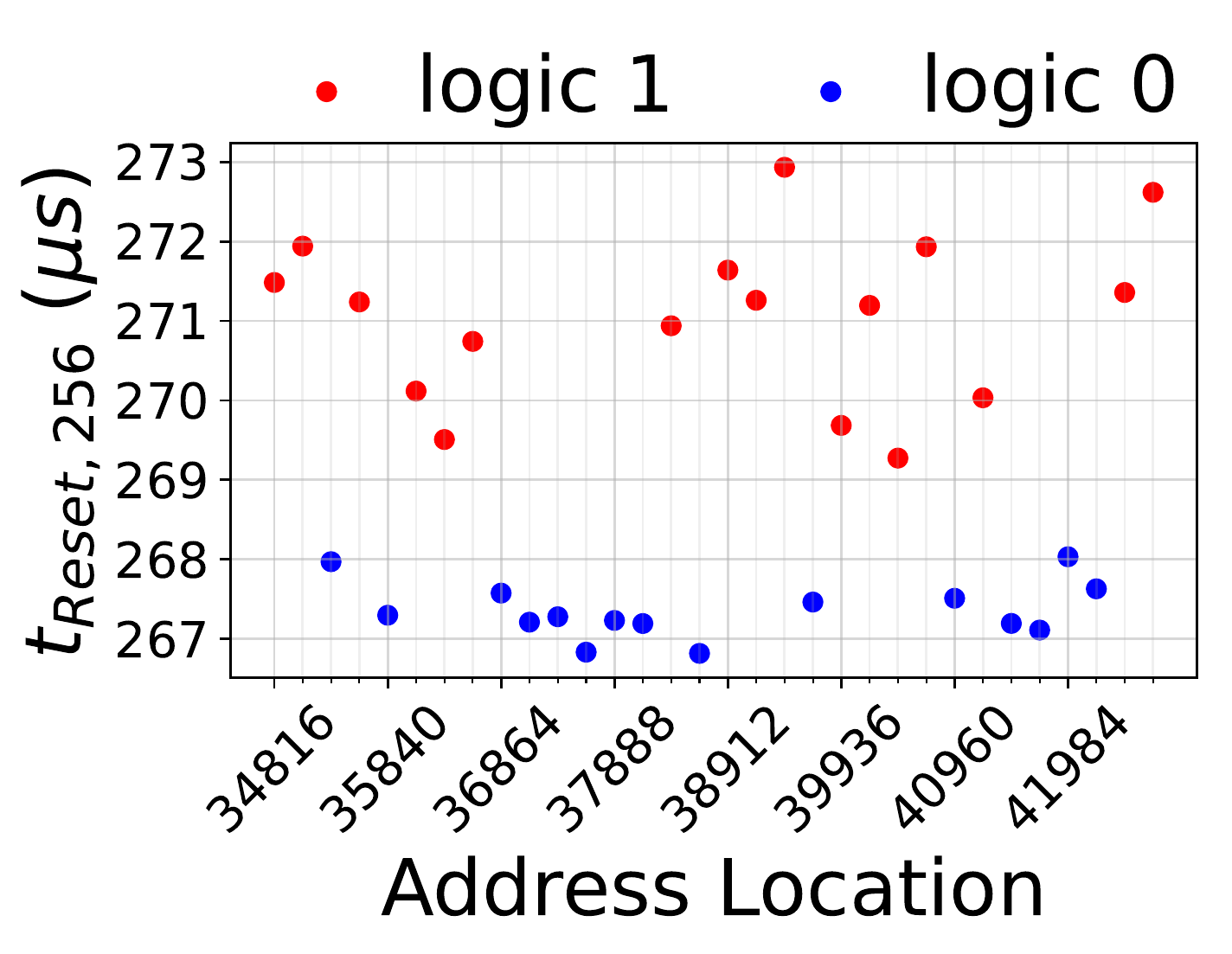}
        \caption{}
        \label{fig:bake_HT_reset}
    \end{subfigure}
    \caption[]{Robustness analysis after high-temperature baking ($80^{\circ}C$) with- (a) $t_{Set,256}$ (b) $t_{Set,256}$}
    \label{fig:robust}
\end{figure}

The watermark should be resilient to the variation of operating conditions, i.e., it will not be possible to modify or change the watermark information with localized heating or operating voltage. Inherently, all modern ICs are resilient to small variations in operating voltage as they are usually integrated with a voltage regulator. Voltage regulators are capable of retaining the operating voltage within a valid range of supply voltage. However, to verify the robustness of our imprinting technique against the temperature, first, we have watermarked a fixed address-space with $15K$ stress. Then we have isolated watermarked memory chip from the system and baked it at $80^{\circ}C$ for $3$ hours. Lastly, we have evaluated the $t_{Set,256}$ and $t_{Reset,256}$ while maintaining the chip temperature of $80^{\circ}C$. We have observed that the watermark information is not affected by temperature and remains well-separated (Fig. \ref{fig:robust}) after the high-temperature baking and high-temperature system-level operation (considering both $t_{Set,256}$ and $t_{Reset,256}$). Such behavior of ReRAM is expected as the resistance ratio of \textit{HRS}/\textit{LRS} is relatively temperature insensitive \cite{Bogdan:ReRAM}.  Note that, ReRAM chips that we have used in our experiment are rated to operate up to $85^{\circ}C$.

%

\subsection{Performance Analysis}\label{subsec:perf}
\subsubsection{Imprinting Time}
The proposed technique for imprinting watermarks relies on repeatedly switching state of ReRAM cells. Thus, the time required to imprint the watermark is directly proportional to the number of stress count, $\mathcal{N}$.
The estimated time to imprint watermark is, $\mathcal{T}_{imprint} = (\mathcal{N} \times \mathcal{B}_{WMark} \times \mathcal{T}_{switch_{pair}})$; where $\mathcal{T}_{switch_{pair}} = (\mathcal{T}_{set} + \mathcal{T}_{reset})$ represents stressing time (\textit{set-reset} pair) for 256 addresses (switching resistance state with single \textit{write} command), and $\mathcal{B}_{WMark}$ represents the number of imprinted bits. The chip used for our experimental evaluation has the following timing parameters: $\mathcal{T}_{switch_{pair}} = (5ms +5ms) = 10ms$, and $\mathcal{B}_{WMark} = 32$. Thus, the baseline implementation requires $((5ms +5ms) \times 32 \times 10k) = 3200s$ for $10K$ switching operations to imprint the watermark. Therefore, the throughput for the watermark imprinting is $\frac{32bits}{3200s}= 0.6bit/min$. 
It is worth mentioning that the imprinting time of our proposed technique heavily depends on the \textit{write} speed of the ReRAM chips. Fortunately, in the past few years, the \textit{write} speed of ReRAM chips significantly improved and will continue to improve in the future. For example, the \textit{write} speed of \textit{MB85AS8MT} ReRAM chips is improved ${>}3X$ over its previous generation \textit{MB85AS4MT} ReRAM chips\footnote{\textit{MB85AS8MT} and \textit{MB85AS4MT} chips were launched in 2019 and 2016, respectively.}.

\subsubsection{Retrieval Time}
Unlike the imprinting procedure, the extraction procedure is significantly fast. The estimated time to retrieve the watermark can be calculated by- $\mathcal{T}_{retrieve} = (\mathcal{T}_{switch} \times \mathcal{B}_{WMark} \times \mathcal{N}_{rep})$; where $\mathcal{T}_{switch}$ is the average value of $t_{Set,256}$ or $t_{Reset,256}$; and $\mathcal{N}_{rep}$ represents the number of addresses used to imprint single bits. After $10K$ stressing, the average value $t_{Set,256}$ is ${\sim}250{\mu}s$, and we used $\mathcal{N}_{rep}=256$ in our implementation. Therefore, the throughput for the watermark retrieval is $\frac{\mathcal{B}_{WMark}}{\mathcal{T}_{retrieve}}=\frac{32bits}{250{\mu}s \times 32 \times 256} = 15.625bits/s$.

\subsubsection{Watermarking Cost}
Our proposed technique only requires $10K$ \textit{set-reset} operations (i.e., $20K$ rewrite cycles) to make a distinguishable separation between logic `0' and `1' of the imprinted watermark (using $t_{Set,256}$). However, the rated endurance of ReRAM chips is $1M$. Therefore, our proposed technique costs only $2\%$ of the rated endurance of imprinted addresses

\section{Conclusion} \label{sec:end}

This paper demonstrated a cost-effective watermark imprinting and extraction technique using commercially available ReRAM chips. In our proposed technique, we utilize repeated switching operations to change the physical properties of the memory cells. The effectiveness of the proposed technique is evaluated by metrics of interest, i.e., the bit separation, imprinting throughput, extraction time, and imprinting cost. Additionally, our proposed technique is robust against temperature variation and does not require any hardware modifications.

\section{Acknowledgments}
This work was supported by the National Science Foundation under Grant Number DGE-2114200. We would also like to thank Mr. Tomohiro Kawakubo of Fujitsu Semiconductor Limited for sharing the necessary ReRAM chip information.

\bibliographystyle{ACM-Reference-Format}
\bibliography{ref}










\end{document}